**Article Type:** Full Paper

# High Conduction Hopping Behavior Induced in Transition Metal Dichalcogenides by Percolating Defect Networks: Toward Atomically Thin Circuits

*Michael G. Stanford, Pushpa R. Pudasaini, Elisabeth T. Gallmeier, Nicholas Cross, Liangbo Liang, Akinola Oyedele, Gerd Duscher, Masoud Mahjouri-Samani, Kai Wang, Kai Xiao, David B. Geohegan, Alex Belianinov, Bobby G. Sumpter, and Philip D. Rack\**

Michael G. Stanford, Dr. Pushpa R. Pudasaini, Nicholas Cross, Prof. Gerd Duscher, Prof. Philip D. Rack

Department of Materials Science and Engineering, University of Tennessee, Knoxville, Tennessee 37996, United States

Email: prack@utk.edu

Elisabeth T. Gallmeier, Dr. Liangbo Liang, Akinola Oyedele, Dr. Masoud Mahjouri-Samani, Dr. Kai Wang, Dr. Kai Xiao, Dr. David B. Geohegan, Dr. Alex Belianinov, Dr. Bobby G. Sumpter, and Dr. Philip D. Rack

Center for Nanophase Materials Sciences, Oak Ridge National Laboratory, Oak Ridge, Tennessee 37831, United States

Dr. Gerd Duscher

Materials Science and Technology Division, Oak Ridge National Laboratory, Oak Ridge, Tennessee 37831, United States

Bobby G. Sumpter

Computational Sciences & Engineering Division, Oak Ridge National Laboratory, Oak Ridge, Tennessee 37831, United States

Keywords: 2D materials, Helium Ion Microscopy, atomically thin circuits, transition metal dichalcogenides, Defect networks

**Abstract**

Atomically thin circuits have recently been explored for applications in next-generation electronics and optoelectronics and have been demonstrated with two-dimensional lateral heterojunctions. In order to form true 2D circuitry from a single material, electronic properties must be spatially tunable. Here, we report tunable transport behavior which was introduced into single layer tungsten diselenide and tungsten disulfide by focused $He^+$ irradiation. Pseudo-metallic behavior was induced by irradiating the materials with a dose of ~$1 \times 10^{16}$ $He^+/cm^2$ to introduce defect states, and subsequent temperature-dependent transport measurements suggest a nearest neighbor hopping mechanism is operative. Scanning transmission electron microscopy and electron energy loss spectroscopy reveal that Se is sputtered preferentially, and extended percolating networks of edge states form within $WSe_2$ at a critical dose of $1 \times 10^{16}$ $He^+/cm^2$. First-principles calculations confirm the semiconductor-to-metallic transition of $WSe_2$ after pore and edge defects were introduced by $He^+$ irradiation. The hopping conduction was utilized to direct-write resistor loaded logic circuits in $WSe_2$ and $WS_2$ with a voltage gain of greater than 5. Edge contacted thin film transistors were also fabricated with a high on/off ratio (> $10^6$), demonstrating potential for the formation of atomically thin circuits.

## 1. Introduction

Transition-metal dichalcogenides (TMDs) are layered materials that consist of stacked two-dimensional crystals held together by van der Waals interactions. An individual TMD layer consists of a X-M-X sandwich structure where X is a chalcogen, and M is a transition-metal atom typically of trigonal prismatic coordination for a 2H polytype[1]. TMDs have been shown to exhibit semiconducting properties and have recently been used to create high performance transistors[2–4] and optical devices[5–7]. Single-layer (1-L) TMDs also exhibit a direct bandgap, typically on the order of 1 - 2 eV depending upon the chemical composition[8].

Defects in TMDs can alter the electronic behavior and band structure in a wide variety of ways. Point defects (0-D), such as vacancies, have been studied extensively for the most commonly synthesized TMDs. For instance, intrinsic chalcogen vacancies in $MoS_2$ exhibit hopping transport characteristics explained by the Mott formalism[9]. Mo vacancies can also induce p-type behavior, whereas S vacancies exhibit n-type behavior[10]. In $WSe_2$ and $MoS_2$, it was shown that chalcogen and transition-metal vacancies introduce mid-gap states into the material[9,11,12], a behavior exhibited in many TMDs[13].

Line defects (1-D), have demonstrated interesting properties with implications on the electrical transport. Different types of 60º grain boundaries, in chemical vapor deposition (CVD) grown TMDs, behave as 1-D metallic wires[11,14]. It is also predicted that the majority of edge termination states behave metallically for $MoS_2$[15,16]. For a review of defect engineering in TMDs see ref [17].

Point defects in TMDs can occur intrinsically; however, a variety of methods have been explored for their control. Nonstoichiometric growth techniques have been used to synthesize

MoSe$_2$ with Se vacancies approaching ~20%[18]. Plasma treatment has the potential to introduce defects as well as to dope TMDs[19–22]. Electron beam irradiation has been used to introduce a variety of point and extended defects[23,24]. More recently, a focused helium ion beam has been used to introduce defects into MoSe$_2$[25], MoS$_2$[26], and WSe$_2$[27], controllably tuning the materials' properties. However, the mechanisms in which He$^+$ irradiation altered electrical properties was not thoroughly explored. Molecular dynamics simulations have demonstrated that defects can be introduced into the lattice of TMDs with a high degree of spatial control by tuning the He$^+$ beam conditions[28].

In this work, a focused He$^+$ beam was used to introduce defects into single layer WSe$_2$. The extent of defects introduction was controlled by adjusting the He$^+$ exposure dose from $1\times10^{14}$ to $1\times10^{17}$ He$^+$/cm$^2$. We found that a dose of $1\times10^{16}$ He$^+$/cm$^2$ induces metallic-like behavior in the WSe$_2$ with transport properties independent of gate voltage. Temperature-dependent transport measurements reveal a thermally activated conductivity mechanism consistent with a nearest-neighbor hopping mechanism. Scanning transmission electron microscopy (STEM) suggests that percolating defect networks are formed at a dose of $1\times10^{16}$ He$^+$/cm$^2$, and Se is preferentially sputtered, resulting in the metallic-like behavior. Density functional theory (DFT) calculations were used to model various defect configurations, and confirm that pore and edge defects induced by the high He$^+$ dose result in many in-gap states in WSe$_2$, and the band gap of the defective system is effectively zero. We demonstrate that the metallic-like WSe$_2$ and WS$_2$ can be utilized to direct-write logic gates on single TMD flakes and direct-write edge contacted transistors. Our results demonstrate a strategy for generating atomically thin circuitry and show potential for large scale processing via standard ion implantation or plasma exposure.

**2. Results and discussion**

Back-gated, 1L, WSe$_2$ field effect transistors (FETs) were fabricated on 290 nm SiO$_2$/Si substrate as described in the Methods section. **Figure 1a** is an optical micrograph of a typical FET device. After fabrication, the channel of the FETs was irradiated with a focused He$^+$ beam at the beam energy of 25 keV. The He$^+$ has shown the potential to introduce defects into 2D materials with a high degree of precision[29–31] by generating vacancies, adatoms, and interstitials within the material lattice via nuclear collisions. First-principles studies of native defects reveal the following: (1) Chalcogen vacancies are the most energetically favored point defect; (2) Chalcogen interstitials experience an extremely low diffusion barrier and are thus expected to rapidly annihilate with vacancy sites; (3) It is energetically favorable for interstitial transition metal atoms to form a split-interstitial configuration[32]. Therefore, it is expected that most point-defects induced by He$^+$ irradiation will be in the form of chalcogen vacancies, transition metal vacancies, or transition metal split-interstitials. Minimal amounts of He are implanted within WSe$_2$ since it is 1L thick, and 25 keV He$^+$ has an interaction volume and peak implantation of hundreds of nanometers in typical 2D materials and substrates. Photoluminescence and Raman spectra as a function of He$^+$ dose are reported in the Supporting Information and demonstrate a reduced PL intensity and quenching of the E$^1_{2g}$ and A$_{1g}$ peaks. **Figure 1b** shows the room temperature transfer curves for FETs in which the channel regions were exposed to varying He$^+$ doses. The pristine WSe$_2$ exhibits p-type behavior with an I$_{on}$/I$_{off}$ of 10$^8$. At a dose of 1×10$^{14}$ He$^+$/cm$^2$, the device on-current degrades as defects are introduced, and I$_{on}$/I$_{off}$ decreases from 10$^8$ to 10$^5$. Furthermore, at a dose of 1×10$^{15}$ He$^+$/cm$^2$, the back gate modulation no longer has an effect on the device current, however at zero gate voltage the current is two orders of magnitude higher than the pristine/un-exposed device. At a dose of 1×10$^{16}$ He$^+$/cm$^2$, the gate modulation is also suppressed, and the zero gate voltage current is approximately 6 orders of magnitude higher than the pristine device. This result is indicative of charge carriers in the absence of a gate applied field. At a dose of 1×10$^{17}$ He$^+$/cm$^2$ and greater, the single layer material is mostly sputtered away, and no current flows. Several devices that demonstrate the same qualitative dose-dependent behavior were measured. **Figure 1c** summarizes the $I_{Exposed}/I_{Pristine}$ ratio for devices exposed to varying doses of $He^+$ at 10 V and -60 V gate voltage, which correspond to the devices "off" and "on" states, respectively. At $V_{GS} = 60$ V, the devices exhibit decreasing current and increasing resistivity as the dose is increased to $1\times10^{15}$ $He^+/cm^2$. The resistivity drops sharply at a dose of $1\times10^{16}$ $He^+/cm^2$, the point at which metallic behavior is exhibited. However, at $V_{GS} = 10$ V, the current increases with increasing $He^+$ dose over the entire range studied. This indicates that carrier concentration is increasing as a result of $He^+$ exposure in the absence of gate induced carrier generation.

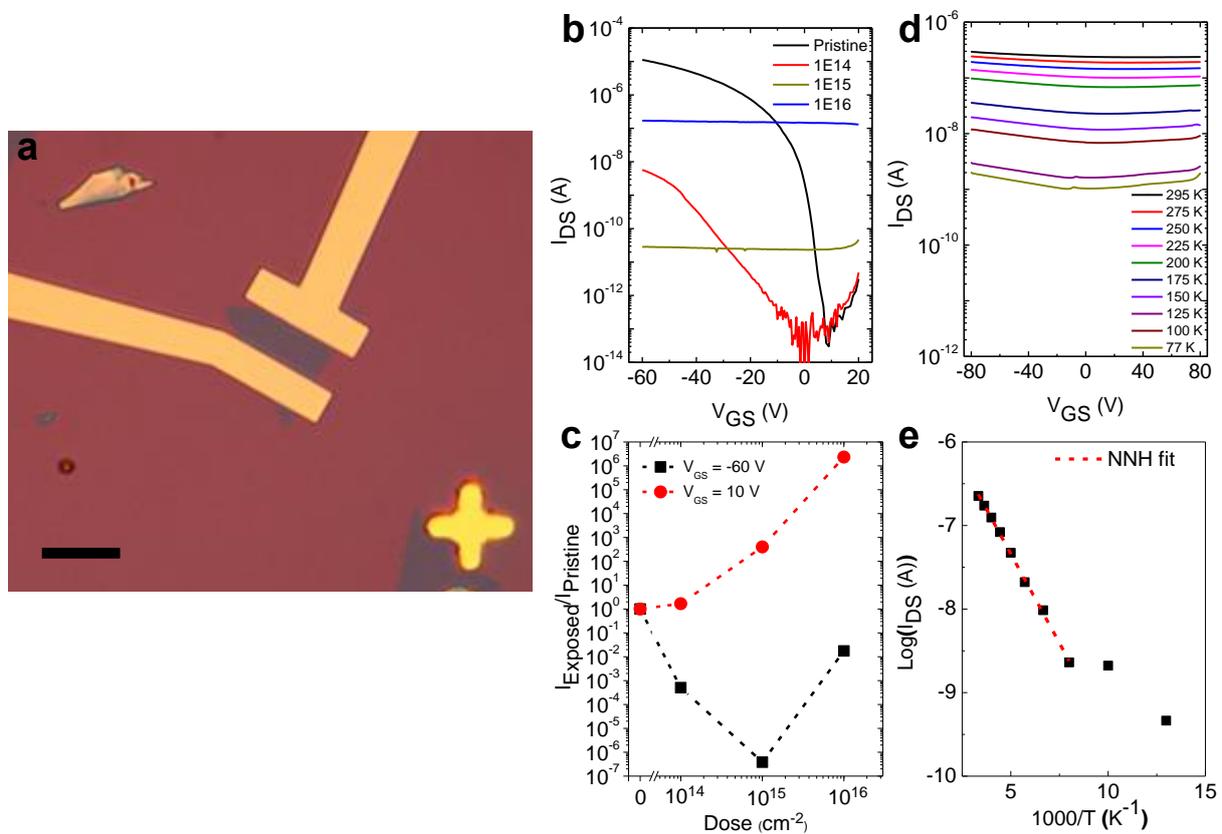

**Figure 1 | $He^+$ dose dependent electrical and optical properties.** (**a**) Optical micrograph of a standard $WSe_2$ FET device. Scale bar is 10 μm. (**b**) Transfer curves for exposed FETs with $V_{DS} = -1.1$ V. (**c**) Plot of the current ratio: current of the exposed device to current of pristine

device at $V_{GS}$ = 10 and -60 V. (**d**) Temperature-dependent transfer curves for WSe$_2$ exposed with a dose of $1\times10^{16}$ He$^+$/cm$^2$. (**e**) Arrhenius plot of the device current ($I_{DS}$) and fitting results of nearest-neighbor hopping model with an activation energy ($E_a$) of 36.7 meV.

To further characterize the conduction mechanism, temperature-dependent transport measurements were conducted on the $1\times10^{16}$ He$^+$/cm$^2$ exposed WSe$_2$ sample. **Figure 1d** shows the transfer curves ranging from room temperature to 77 K. The device on-current is largely independent of gate modulation for all reported temperatures, and there is a sharp decrease in the conductivity as the temperature decreases. This deviates from the standard metallic behavior where phonon-scattering is reduced at low temperature, thereby increasing conductivity. The temperature-dependent source-drain I-V characteristics of the device exhibit linear behavior and are reported in the Supporting Information. The sheet resistance ($R_s$) at room temperature is approximately $7.2\times10^6$ Ω/sq and increases to $3.2\times10^9$ Ω/sq at 77 K; For reference, the $R_s$ for graphene has a range from $10^2 - 10^6$ Ω/sq[33] at room temperature depending upon the synthesis technique. **Figure 1e** is the Arrhenius plot of the device conductance (G = I/V) at the measured temperatures. The behavior can be broken up into a high-temperature (295-125 K) and a low-temperature regime (below 125 K). At high temperature, the transport exhibits an activation behavior associated with a nearest-neighbor hopping (NNH) transport mechanism for which the current is governed by Equation 1:

$$G \propto \exp(-\frac{E_a}{k_b T}) \qquad \text{Eq. (1)}$$

The activation energy ($E_a$) was determined to be ~ 36.7 meV, a reasonable value for hopping transport reported in other 2D materials[34]. Below a characteristic crossover temperature ($T_c$) of 125 K, the transport deviates from the Arrhenius behavior. At low temperature, variable range hopping (VRH), which has a T$^{1/3}$ dependence, appears to dominate the transport mechanism over NNH. Due to the temperature range studied, we cannot accurately fit this portion of the plot. The VRH – NNH crossover transport is commonly associated with disordered semiconductors[35], and has been previously observed in TMDs with intrinsic point defects[9]. However, the observed behavior differs from other reports due to the independence from gate modulation.

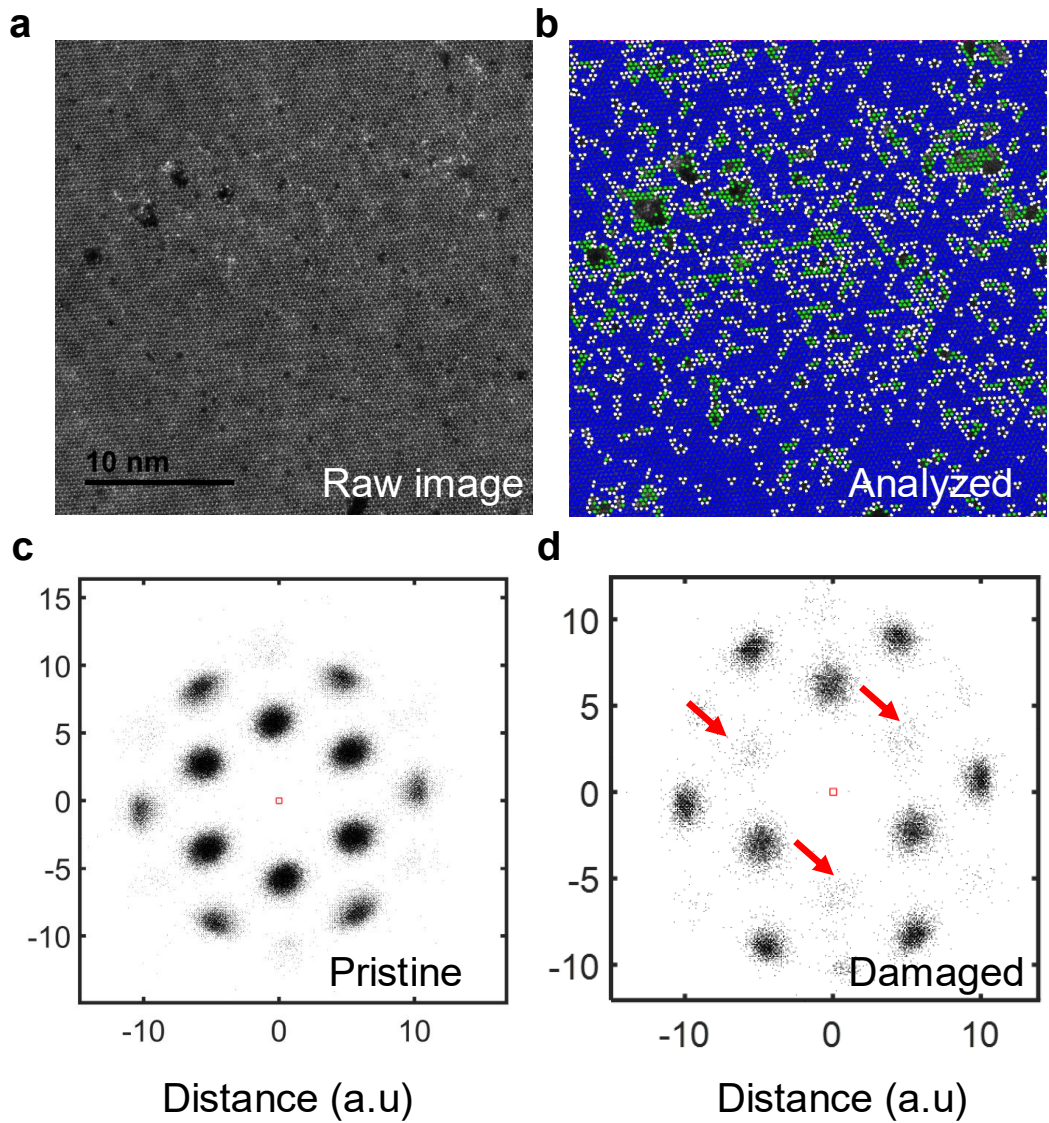

**Figure 2 | STEM image of exposed WSe$_2$.** (**a**) HAADF STEM image of WSe$_2$ which was exposed with a dose of $1 \times 10^{15}$ He$^+$/cm$^2$. Field of view is 32 nm. (**b**) STEM image local crystallography analysis using K-means clustering, on the six-nearest neighbor bond-lengths to each atom. Blue atoms represent a pristine lattice, white atoms are around point defects (light damage), and green atoms are extended defects (heavy damage). Nearest neighbor distribution

in real space for (**c**) pristine (blue) and (**d**) lightly damaged (white) atoms in the lattice. Lightly damage cluster shows significantly lower counts for neighbors in 'downward pointing' trimer, which corresponds with preferentially sputtered Se.

To gain a better understanding of the origin of the He$^+$ dose dependent transport behavior in the 1L WSe$_2$, STEM images of exposed WSe$_2$ were analyzed to reveal signatures of preferential Se sputtering. **Figure 2a** shows a STEM image of WSe$_2$, which was exposed with a dose of $1\times10^{15}$ He$^+$/cm$^2$. A dose of $1\times10^{15}$ He$^+$/cm$^2$ is reported here because He$^+$-induced defects can easily be distinguished, although the WSe$_2$ is still continuous. K-means cluster analysis, using a Euclidean metric on bond-lengths to a six-member neighborhood for each atom, was conducted on the STEM image and is reported in **Figure 2b.** This approach clusters each atom in the image based on the bond-length to six nearest neighbors (see details in Methods section[36–38]). This analysis allows automated detection of atoms in varying crystallographic configurations. For instance, pristine lattice atoms with a consistent periodic configuration are labeled blue; atoms neighboring single vacancies are labeled white; and atoms from heavily damaged regions with multiple neighboring vacancies are labeled green. **Figure 2c** shows the real space nearest neighbor distribution for all atoms in the image characterized as pristine (blue) by the k-means analysis. Since the lattice is pristine, all neighboring positions are equally occupied, as indicated by their similar contrast at each point of the hexagon closest to the center of the image. The real space nearest neighbor distribution for atoms neighboring single vacancies (green) are reported in **Figure 2d**. Nearest neighbor atoms that occupied the downward facing trimer (indicated by inset red arrows) have fewer counts, as indicated by their lighter contrast. This suggests that atoms occupying these sites were preferentially sputtered by the He$^+$ irradiation in comparison to the other nearest neighbors. Image analysis reveals that the preferentially sputtered atomic species are Se atoms. Therefore, He$^+$ irradiation results in chalcogen deficiency in irradiated films although some metal atoms are also sputtered. Electron energy loss spectroscopy (EELS) analysis confirms the He$^+$ irradiation creates Se deficiency within the flakes (see Supporting Information Figure S4).

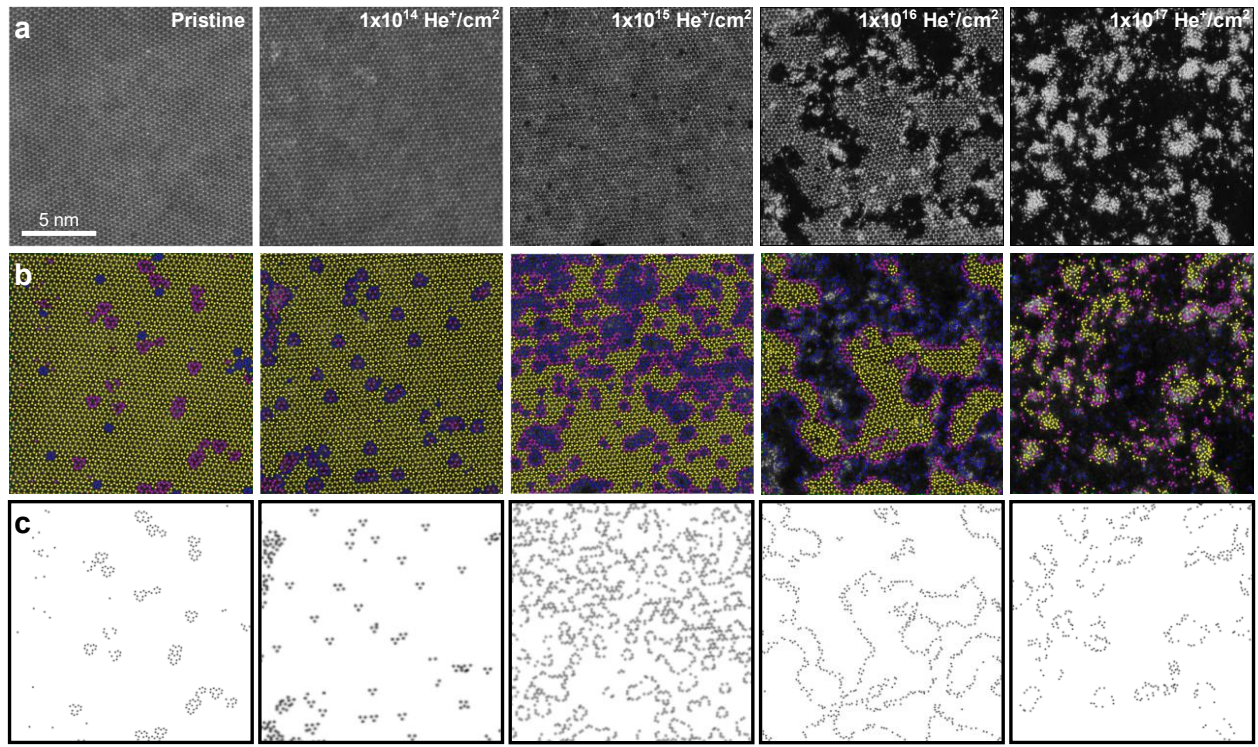

**Figure 3 | STEM images and analysis.** (**a**) High angle annular dark field (HAADF) STEM images of WSe$_2$ irradiated with varying doses of He$^+$ (text inset in image). (**b**) K-means cluster analyzed STEM images. (**c**) Only the vacancy boundary atoms (magenta color in the mid row) from the STEM images (top row).

STEM images of the WSe$_2$ were collected as a function of dose and reported in **Figure 3a** (doses are listed as insets). He$^+$ irradiation introduces more structural defects into the lattice as a function of increasing dose. At $1\times10^{14}$ He$^+$/cm$^2$, isolated, predominately Se vacancies are generated, which leads to carrier scattering and reduces the field effect mobility from ~ 16 cm$^2$/Vs for a pristine device to ~$1.5\times10^{-2}$ cm$^2$/Vs (Details of field effect mobility calculation can be found in the methods). Notably, at a dose of $1\times10^{15}$ He$^+$/cm$^2$, the Se vacancy concentration increases to a point that the gate-independent NNH transport dominates the conduction relative to the p-type semiconducting behavior of the pristine and $1\times10^{14}$ He$^+$/cm$^2$ samples. At a dose of $1\times10^{16}$ He$^+$/cm$^2$, the defect density is an order of magnitude higher and defects begin coalescing into an extended network percolating throughout the material. Consequently, a greater density of Se vacancies lowers the activation energy and the hopping distance, thereby contributing to an increase in conductivity. A dose of $1\times10^{17}$ He$^+$/cm$^2$ sputters much of the material leaving discontinuous patches. K-means cluster analysis results are reported in **Figure 3b.** Pristine lattice atoms with a periodic configuration of neighboring atoms are labeled as yellow; atoms neighboring vacancies (boundary atoms) are labeled as magenta; and atoms with little or no detectable periodicity are labeled as blue. The boundary states are isolated and reported in **Figure 3c** for clarity**.** At a dose of $1\times10^{14}$ - $1\times10^{15}$ He$^+$/cm$^2$, there are numerous boundary atoms surrounding pores and point defects, but they are largely isolated from each other. However, at a dose of $1\times10^{16}$ He$^+$/cm$^2$, the boundary atoms form an extended percolating network in the single layer WSe$_2$. It is worth noting that the defect concentration in the free-standing WSe$_2$ induced by a particular He$^+$ dose is slightly lower than the defect concentration that can be expected for material exposed on a bulk substrate. This is because the primary He$^+$ beam energy has a small interaction cross-section with the freestanding WSe$_2$[28] and back-scattered He$^+$ and recoil atoms contribute significantly toward defect introduction. For atomically thin free-standing material, the vast majority of He$^+$ transmit or forward-scatter through the material, thus minimizing defects generated from backscattered He$^+$. However, recoil atom cascades are still expected to significantly contribute toward defect production. For a more detailed description on how the underlying substrate effects the He$^+$-induced defect introduction, interested readers are referred to Ref [39].

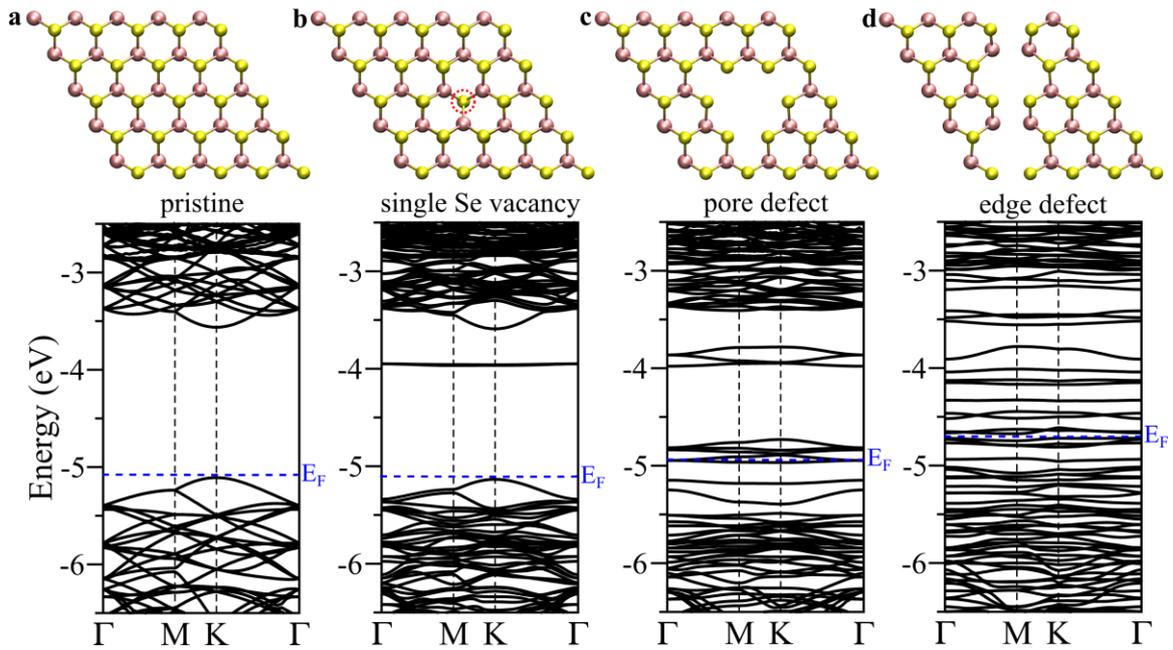

**Figure 4 | DFT results for pristine and defective single layer WSe$_2$.** Calculated electronic band structures of the 5×5 supercell with (**a**) no defect, (**b**) a point defect, (**c**) a pore defect, and (**d**) edge defect configurations. All band energies are aligned to the vacuum potential for direct comparison. The Fermi level is set at the valence band maximum for each system, as shown by the blue dashed line. The corresponding atomic structure for each configuration is shown above the band structure. The point defect in (b) corresponds to a single Se vacancy, which is indicated by the red dashed circle.

Unfortunately, this image analysis does not reveal a single dominant configuration for the edge termination in the percolating network. In context with previous reports, edge states behave metallically for most commonly occurring edge terminations[15,16]. Therefore, percolating networks of edge states surrounding pores are believed to behave as conductive nanowires and contribute to the high conductivity at $1\times10^{16}$ He$^+$/cm$^2$. To confirm our analysis, first-principles DFT calculations were performed for different defect configurations, as shown in **Figure 4**. Pristine monolayer WSe$_2$ has a direct band gap of ~1.55 eV at the K point, according to this level of DFT calculations. For the point defect which is predominately a Se vacancy induced by a low irradiation dose such as $1\times10^{14}$ He$^+$/cm$^2$, this leads to two nearly degenerate in-gap bands above the Fermi level, which are almost dispersionless as their charge densities are well localized around the point defect site (**Figure 4b**)[9,25]. Thus the point defect lowers the carrier mobility and also the band gap to ~1.17 eV. Although it retains the semiconducting feature, the gap reduction indicates the decreased current on/off ratio, consistent with our experimental observation. With increasing irradiation dose, pore and edge defects are introduced according to the STEM images, giving rise to more in-gap bands. Our calculations suggest that many in-gap states are close to the Fermi level, effectively closing the band gap and rendering the system metallic. This can explain why the back gate modulation no longer has an effect on the device current at doses of $1\times10^{15}$ and $1\times10^{16}$ He$^+$/cm$^2$, and the zero gate voltage current is much higher than that of the pristine device. In other words, both the circular edge surrounding the pore defect (**Figure 4c**) and the straight edge (**Figure 4d**) lead to metallic edge states that behave as conductive nanowires. However, isolated pores, which are prevalent with exposure to a dose of $1\times10^{15}$ He$^+$/cm$^2$, do not coalesce into extended networks and result in lower conduction than material exposed to a dose of $1\times10^{16}$ He$^+$/cm$^2$. Thus, conduction in the material results from percolating 1-D networks of edge states in series with pockets of Se deficient material. The NNH temperature dependence is a result of the Se deficient material which acts as a bottleneck in the conduction path between high conductivity edge states, as NNH behavior has been demonstrated in TMDs with intrinsic point defects[9].

Since gate modulation has a minimal effect on the carrier concentration of the percolating defect networks, the NNH WSe$_2$ can be utilized to direct write circuitry and logic gates onto a single flake. **Figure 5a** shows a proposed schematic for the structure of a resistor-loaded inverter device (circuit diagram is inset in **Figure 5c**). This logic gate is fabricated as a series of two devices. The channel between the first two electrodes is irradiated with a dose of $1\times10^{16}$ He$^+$/cm$^2$ in order to induce gate-independent NNH behavior. This region behaves as a resistor due to the linear I-V characteristics. The channel between the second and third electrodes is pristine WSe$_2$. This device behaves as a standard p-FET with significant gate modulation. An optical micrograph of a fabricated invertor is displayed in **Figure 5b.** The input (V$_{in}$) – output (V$_{out}$) characteristics for a back-gated (SiO$_2$) WSe$_2$ invertor are shown in **Figure 5c**, and typical invertor behavior of a p-type device is exhibited. In this device, the NNH channel gives a resistance of approximately 5 MΩ. Transfer curves and I-V characteristics of the transistor and resistor in the invertor structure are reported in the Supporting Information **Figures S5 and S6**. In order to demonstrate the applicability of the NNH defect network in other TMDs, a similar resistor-loaded invertor was fabricated from single layer WS$_2$. Input-output characteristics are shown in **Figure 5d**; notably, this device shows opposite input-output characteristics since the WS$_2$ transistor exhibited n-type behavior. Similar to the WSe$_2$ invertor, the resistor was direct-written with He$^+$ irradiation into a single flake by inducing the NNH defect network. The voltage gain (dV$_{out}$/dV$_{in}$) in these devices is low since they were back-gated with relatively low-dielectric constant SiO$_2$ ($\varepsilon_r$ = 3.9). In order to improve invertor performance, a WSe$_2$ invertor was top-gated with an ionic liquid (IL) (see supporting information for details) and is reported in **Figure 5e**. V$_{DD}$ was varied from 0.5 V to 1.5 V, and V$_{out}$ was consistently at the level of V$_{DD}$ for positive V$_{in}$, when the transistor was in an off state. As V$_{in}$ is decreased to a threshold of -1 V, V$_{out}$ rapidly switches to 0 V and displays ideal invertor characteristics. **Figure 5f** reports the voltage gain, which was greater than 5 for V$_{DD}$ equal to 1.5 V. This is a suitable value for implementation of these devices into electronic circuits where a gain greater than 1 is typically desired to drive the input of the next invertor in a circuit.

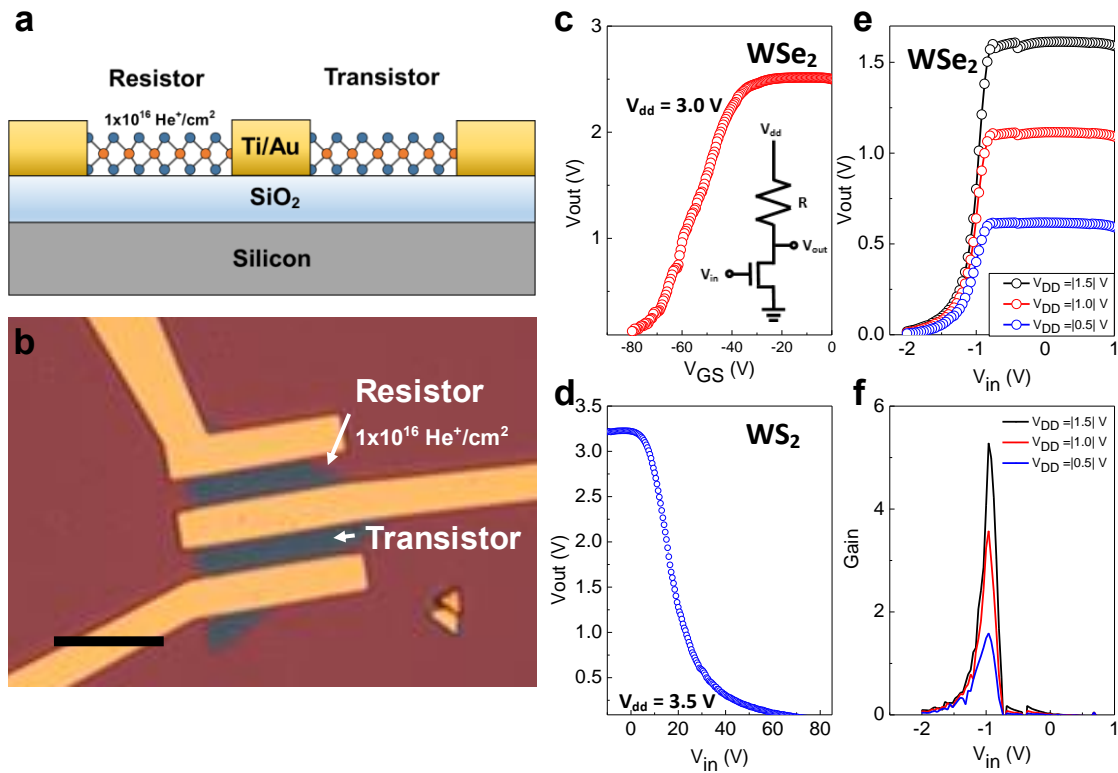

**Figure 5 | Atomic layer inverters.** (**a**) Schematic of atomic layer inverter created by utilizing exposed and pristine material on a single flake. (**b**) Optical micrograph of a typical inverter device. Scale bar is 10 μm. (**c**) Input ($V_{in}$) – Output ($V_{out}$) voltage characteristics of a $WS_2$ atomic layer inverter. Circuit diagram for device is inset. (**d**) Input ($V_{in}$) – Output ($V_{out}$) voltage characteristics of a $WSe_2$ atomic layer inverter. (**e**) Input ($V_{in}$) – Output ($V_{out}$) voltage characteristics of a $WSe_2$ atomic layer inverter which was gated with an ionic liquid. (**f**) Voltage gain ($dV_{out}/dV_{in}$) of inverter reported in figure (**e**).

Furthermore, high conduction NNH behavior induced in the TMD flake can be utilized to create edge contacts for true 2D circuitry, which typically involves a CVD step with a seeding mechanism[40,41]. **Figure 6a** displays a schematic of an edge contacted transistor fabricated from NNH $WS_2$ acting as the source and drain along with pristine $WS_2$, which acts as the channel material. The Raman mapping of the $WS_2$ 2LA(M) peak in **Figure 6b** shows that a relatively sharp transition can be achieved between defective and pristine material. The

2LA(M) peak is clearly suppressed for the defective WS$_2$. The transfer curve of the edge contacted device compared to that of a standard WS$_2$ FET with a Ti/Au source-drain is shown in **Figure 6c**. The introduction of NNH WS$_2$ edge contact results in a slight reduction in the on-current of the device; however, $I_{on}/I_{off}$ remain greater than $10^6$. Total resistance of a device in its on-state, defined as intrinsic resistance plus contact resistance, increases from ~3.7 MΩ to ~14.4 MΩ after the addition of edge contacts. This is likely due to the intrinsic resistivity of the NNH WS$_2$ and could be reduced by further optimizing the defect concentration. Hence, the induced NNH behavior in TMD flakes can serve as a method to write circuitry and logic devices into the material for atomically thin circuitry and edge contacted devices. The high conduction behavior induced by the formation of extended defect networks can likely be scaled-up via standard lithography of a blocking mask coupled with standard ion implantation or plasma processing.

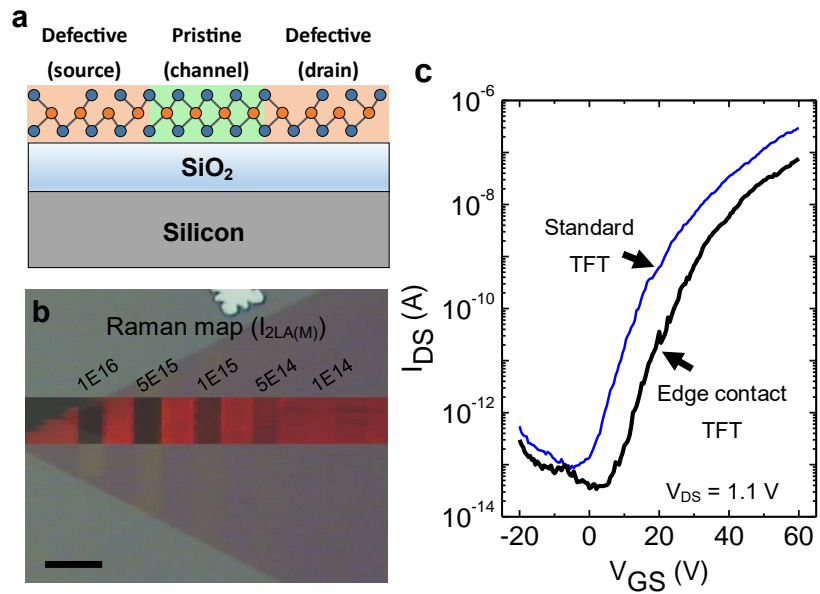

**Figure 6 | Edge contacts for 2D devices.** (**a**) Schematic of edge contact FET. Source and drain are fabricated from NNH WS$_2$ (exposed to a dose of $1\times10^{16}$ He$^+$/cm$^2$). The FET channel is pristine (unexposed) WS$_2$. (**b**) Raman map of a WS$_2$ flake plotting the intensity of the 2LA(M) peak. The WS$_2$ flake was exposed to doses varying from $1\times10^{14} - 1\times10^{16}$ He$^+$/cm$^2$ (inset). Scale

bar is 10 μm. Raman map is overlaid on an optical micrograph. (**c**) Transfer curve for a standard WS$_2$ FET compared with that of an edge contacted FET in which the source and drain were composed of NNH WS$_2$.

## 3. Conclusions

We have demonstrated that extended networks of defects in WSe$_2$ and WS$_2$ can be utilized to induce high conductivity behavior. It is believed, from STEM image analysis, that this behavior arises from the Se deficiency and the formation of extended edge states that percolate throughout the material and behave as 1-D metallic nanowires. DFT simulations confirm that the metallic edge states can be introduced after irradiation. However, temperature-dependent electrical transport suggests that a nearest neighbor hopping transport mechanism is operative within the material, which deviates from standard metallic behavior. The NNH behavior is exhibited due to pockets of chalcogen deficient material which remains between metallic edge states. Networks of defects were induced within single flakes to direct-write inverter devices from WSe$_2$ and WS$_2$, devices that demonstrate a voltage gain suitable for electronic circuitry. The NNH material was also utilized to create edge contacts for a transistor that demonstrated an on-off ratio of greater than $10^6$. This enables the formation of atomically thin circuitry without the need for seeding and a CVD step.

## 4. Experimental Section

### 4.1 Material synthesis

Monolayer WSe$_2$ crystals were synthesized on SiO$_2$/Si substrates by a digital transfer growth method that combines pulsed laser deposition (PLD) and VTG processes[42]. We employ PLD to first deposit a uniform and precise amount of stoichiometric precursor nanoparticles onto a "source" substrate at room temperature. Specifically, SiO$_2$/Si source substrates were placed d = 5cm away and parallel to the WSe$_2$ targets in a vacuum chamber. The WSe$_2$ targets were prepared by pressing the stoichiometric powder and were ablated by an excimer laser (KrF 248 nm, 20 ns, 1 Jcm$^{-2}$) at 30º angle of incidence with a repetition rate of 1 Hz at the background gas pressure 200 mTorr. This material was then covered by a "receiver" substrate which is placed in contact and on top of the source substrate to form a confined growth system. The source substrate was placed in contact with a button heater at about 900 ºC to evaporate the precursor materials on the source substrate. By controlling the background gas pressure, a temperature gradient was established that resulted in condensation of the evaporated precursor materials onto a receiver substrate and growth of WSe$_2$ monolayer crystals.

High-density WS$_2$ monolayer crystals were directly grown on SiO$_2$/Si substrates by chemical vapor deposition. The details were reported in a previous publication[43].

**4.2 Helium ion irradiation**

He$^+$ exposures were conducted in a Zeiss Orion NanoFab microscope. A constant beam energy of 25 keV was used for all exposures. All exposure patterns were generated using the NanoPatterning and Visualization Engine (NPVE) pattern generator produced by Fibics Inc. TMD field-effect transistors were irradiated after device fabrication. For STEM imaging, the WSe$_2$ was transferred to a QUANTIFOIL holey carbon grid prior to He$^+$ irradiation.

**4.3 Raman and Photoluminescence**

Raman spectroscopy and photoluminescence measurements were performed in a Renishaw inVia micro-Raman system using a 532 nm excitation laser. A 100 × magnification objective was used for spectral acquisition with a 10 s acquisition time, and three acquisitions were averaged together. The laser spot size was approximately 0.6 µm. Data analysis was conducted with the WIRE v3.4 software.

**4.4 Device fabrication**

Single-layer WSe$_2$ flakes were transferred onto SiO$_2$ (290 nm)/Si (heavily doped Si, which also serves as a bottom gate electrode). Standard e-beam lithography, followed by e-beam evaporation, was employed to create the source/drain electrodes for electrical measurements. The contacts consisted of Ti/Au (5/30 nm) metals deposited and subsequently patterned via a lift-off process. The electrical characteristics of the fabricated WSe$_2$ devices were measured using an Agilent semiconductor parametric analyzer (Agilent Tech B1500 A). Field effect mobility was calculated using the following equation, $\mu = (L/W)*(1/C_g V_{DS})*(dI_{DS}/dV_{GS})$, where L is the channel length, W is the channel width, $C_g$ is the gate capacitance (~12 nF/cm$^2$), $V_{DS}$ is 1.1 V, and $dI_{DS}/dV_{GS}$ is the slope from the linear portion of the $I_{DS}$ vs $V_{GS}$ curve.

### 4.5 Scanning transmission electron microscope imaging

All STEM microscopy was carried out using a Nion UltraSTEM 100 operating at an accelerating voltage of 100 kV. Flakes of WSe$_2$ were exfoliated from a bulk crystal onto films on continuous α-carbon film grids. The 1L WSe$_2$ images consist of averaging 20 frames each 512 x 512 pixels with a 4 μs dwell time per pixel. The EELS spectra were acquired with a Gatan Spectrometer "Enfina" fitted to the STEM.

### 4.6 STEM image K-means cluster analysis

Quantification of the STEM image data was broken down using the following workflow. First, each atom was located in the image using an atom finding algorithm previously described in Ref 35. Second, vectors describing local 6 member neighborhoods for each atom, based on the six closest atoms, were constructed. These distance vectors served as input into a k-means cluster algorithm, utilizing a Euclidian distance metric, with 1000 iterations to confirm minima convergence. Hierarchical cluster trees (dendrograms) were used to assess the most appropriate number of clusters, and ranges of 2-8 clusters were calculated and analyzed for each image.

### 4.7 Density functional theory

Plane-wave DFT calculations were performed using the VASP package[44] equipped with the projector augmented-wave (PAW) pseudopotentials with a cut-off energy of 300 eV. The exchange-correlation interactions were considered in the generalized gradient approximation (GGA) using the Perdew-Burke-Ernzerhof (PBE) functional[45]. Monolayer $WSe_2$ was modeled by a periodic slab geometry with 18 Å vacuum separation used in the out-of-plane direction to avoid spurious interactions with periodic images. For the hexagonal unit cell of pristine $WSe_2$, its optimized in-plane lattice constant is 3.32 Å and 24×24×1 k-point samplings were used in the Monkhorst-Pack scheme. To model the electronic properties of different defect configurations, the point Se vacancy, pores defect and edge defect were considered in a 5×5 supercell with 6×6×1 k-point samplings used. All atoms were relaxed until the residual forces below 0.01 eV/Å.

**Supporting Information**

Supporting Information is available from the Wiley Online Library or from the author.

**Acknowledgements**

P.D.R., M.G.S., and P.R.P. acknowledge support by US Department of Energy (DOE) under Grant No. DOE DE-SC0002136. N.C. and G.D. acknowledge support from the National Science Foundation (NSF) under Grant No. DMR-1410940. Electron microscopy at ORNL was sponsored by U.S. Department of Energy, Office of Science, Basic Energy Sciences, Materials Sciences and Engineering Division. Authors acknowledge Dr. Ilia N. Ivanov for technical assistance in confocal micro Raman experiments which were conducted under user research proposal CNMS 2016-066. We acknowledge that the device fabrication, Raman spectroscopy, STEM imaging, and helium ion irradiation were conducted at the Center for Nanophase Materials Sciences, which is a DOE Office of Science User Facility. E.T.G., A.B.,

M.M.S., K.W., D.B.G., B.G.S., L.L., and K.X acknowledge support from the Center for Nanophase Materials Sciences. The synthesis of 2D materials (M.M.S, K.W., K.X., D.G.) was supported by the U.S. Department of Energy, Office of Science, Basic Energy Sciences, Materials Sciences and Engineering Division and performed in part as a user project at the Center for Nanophase Materials Sciences.

**References**

[1]   C. Sourisseau, F. Cruege, M. Fouassier, M. Alba, *Chem. Phys.* **1991**, *150*, 281.

[2]   B. Radisavljevic, A. Radenovic, J. Brivio, V. Giacometti, A. Kis, *Nat. Nanotechnol.* **2011**, *6*, 147.

[3]   W. Liu, J. Kang, D. Sarkar, Y. Khatami, D. Jena, K. Banerjee, *Nano Lett.* **2013**, *13*, 1983.

[4]   Y. Yi, C. Wu, H. Liu, J. Zeng, H. He, J. Wang, *Nanoscale* **2015**, *7*, 15711.

[5]   B. W. H. Baugher, H. O. H. Churchill, Y. Yang, P. Jarillo-Herrero, *Nat Nano* **2014**, *9*, 262.

[6]   J. S. Ross, P. Klement, A. M. Jones, N. J. Ghimire, J. Yan, M. G., T. Taniguchi, K. Watanabe, K. Kitamura, W. Yao, D. H. Cobden, X. Xu, *Nat Nano* **2014**, *9*, 268.

[7]   Q. H. Wang, K. Kalantar-Zadeh, A. Kis, J. N. Coleman, M. S. Strano, *Nat. Nanotechnol.* **2012**, *7*, 699.

[8]   G. H. Yousefi, *Mater. Lett.* **1989**, *9*, 38.

[9]   H. Qiu, T. Xu, Z. Wang, W. Ren, H. Nan, Z. Ni, Q. Chen, S. Yuan, F. Miao, F. Song, *Nat. Commun.* **2013**, *4*, 2642.

[10]  S. McDonnell, R. Addou, C. Buie, R. M. Wallace, C. L. Hinkle, *ACS Nano* **2014**, *8*,

2880.

[11] W. Zhou, X. Zou, S. Najmaei, Z. Liu, Y. Shi, J. Kong, J. Lou, P. M. Ajayan, B. I. Yakobson, J.-C. Idrobo, *Nano Lett.* **2013**, *13*, 2615.

[12] J. Lu, A. Carvalho, X. K. Chan, H. Liu, B. Liu, E. S. Tok, K. P. Loh, A. H. Castro Neto, C. H. Sow, *Nano Lett.* **2015**, *15*, 3524.

[13] M. Pandey, F. A. Rasmussen, K. Kuhar, T. Olsen, K. W. Jacobsen, K. S. Thygesen, *Nano Lett.* **2016**, *16*, 2234.

[14] H.-P. Komsa, A. V. Krasheninnikov, *Adv. Electron. Mater.* **2017**, 1600468.

[15] M. V. Bollinger, J. V. Lauritsen, K. W. Jacobsen, J. K. Nørskov, S. Helveg, F. Besenbacher, *Phys. Rev. Lett.* **2001**, *87*, 196803.

[16] M. V. Bollinger, K. W. Jacobsen, J. K. Nørskov, *Phys. Rev. B* **2003**, *67*, 85410.

[17] Z. Lin, B. R. Carvalho, E. Kahn, R. Lv, R. Rao, H. Terrones, M. A. Pimenta, M. Terrones, *2D Mater.* **2016**, *3*, 22002.

[18] M. Mahjouri-Samani, L. Liang, A. Oyedele, Y.-S. Kim, M. Tian, N. Cross, K. Wang, M.-W. Lin, A. Boulesbaa, C. M. Rouleau, A. A. Puretzky, K. Xiao, M. Yoon, G. Eres, G. Duscher, B. G. Sumpter, D. B. Geohegan, *Nano Lett.* **2016**, *16*, 5213.

[19] N. Choudhary, M. R. Islam, N. Kang, L. Tetard, Y. Jung, S. I. Khondaker, *J. Phys. Condens. Matter* **2016**, *28*, 364002.

[20] M. Chen, H. Nam, S. Wi, L. Ji, X. Ren, L. Bian, S. Lu, X. Liang, *Appl. Phys. Lett.* **2013**, *103*, 142110.

[21] A. Azcatl, X. Qin, A. Prakash, C. Zhang, L. Cheng, Q. Wang, N. Lu, M. J. Kim, J. Kim, K. Cho, R. Addou, C. L. Hinkle, J. Appenzeller, R. M. Wallace, *Nano Lett.* **2016**,

*16*, 5437.

[22] S. I. Khondaker, M. R. Islam, *J. Phys. Chem. C* **2016**, *120*, 13801.

[23] H.-P. Komsa, S. Kurasch, O. Lehtinen, U. Kaiser, A. V Krasheninnikov, *Phys. Rev. B* **2013**, *88*.

[24] H.-P. Komsa, J. Kotakoski, S. Kurasch, O. Lehtinen, U. Kaiser, A. V. Krasheninnikov, *Phys. Rev. Lett.* **2012**, *109*, 35503.

[25] V. Iberi, L. Liang, A. V. Ievlev, M. G. Stanford, M.-W. Lin, X. Li, M. Mahjouri-Samani, S. Jesse, B. G. Sumpter, S. V. Kalinin, D. C. Joy, K. Xiao, A. Belianinov, O. S. Ovchinnikova, *Sci. Rep.* **2016**, *6*, 30481.

[26] D. S. Fox, Y. Zhou, P. Maguire, A. O'Neill, C. Ó'Coileáin, R. Gatensby, A. M. Glushenkov, T. Tao, G. S. Duesberg, I. V Shvets, M. Abid, M. Abid, H.-C. Wu, Y. Chen, J. N. Coleman, J. F. Donegan, H. Zhang, *Nano Lett.* **2015**, *15*, 5307.

[27] M. G. Stanford, P. R. Pudasaini, A. Belianinov, N. Cross, J. H. Noh, M. R. Koehler, D. G. Mandrus, G. Duscher, A. J. Rondinone, I. N. Ivanov, T. Z. Ward, P. D. Rack, *Sci. Rep.* **2016**, *6*, 27276.

[28] M. Ghorbani-Asl, S. Kretschmer, D. E. Spearot, A. V Krasheninnikov, *2D Mater.* **2017**, *4*, 25078.

[29] J. Buchheim, R. M. Wyss, I. Shorubalko, H. G. Park, *Nanoscale* **2016**, *8*, 8345.

[30] S. Hang, Z. Moktadir, H. Mizuta, *Carbon N. Y.* **2014**, *72*, 233.

[31] K. Yoon, A. Rahnamoun, J. L. Swett, V. Iberi, D. A. Cullen, I. V. Vlassiouk, A. Belianinov, S. Jesse, X. Sang, O. S. Ovchinnikova, A. J. Rondinone, R. R. Unocic, A. C. T. van Duin, *ACS Nano* **2016**, *10*, 8376.

[32] H.-P. Komsa, A. V. Krasheninnikov, *Phys. Rev. B* **2015**, *91*, 125304.

[33] S. De, J. N. Coleman, *ACS Nano* **2010**, *4*, 2713.

[34] M. Y. Han, J. C. Brant, P. Kim, *Phys. Rev. Lett.* **2010**, *104*, 56801.

[35] A. Yildiz, N. Serin, T. Serin, M. Kasap, *Jpn. J. Appl. Phys.* **2009**, *48*, 111203.

[36] A. Belianinov, Q. He, M. Kravchenko, S. Jesse, A. Borisevich, S. V. Kalinin, *Nat. Commun.* **2015**, *6*, 7801.

[37] Q. He, J. Woo, A. Belianinov, V. V. Guliants, A. Y. Borisevich, *ACS Nano* **2015**, *9*, 3470.

[38] K. Yoon, A. Rahnamoun, J. L. Swett, V. Iberi, D. A. Cullen, I. V. Vlassiouk, A. Belianinov, S. Jesse, X. Sang, O. S. Ovchinnikova, A. J. Rondinone, R. R. Unocic, A. C. T. van Duin, *ACS Nano* **2016**, *10*, 8376.

[39] M. G. Stanford, P. R. Pudasaini, N. Cross, K. Mahady, A. Hoffman, D. G. Mandrus, G. Duscher, M. F. Chisholm, P. D. Rack, *Small Methods* **2017**, *1*, 1600060.

[40] M. Zhao, Y. Ye, Y. Han, Y. Xia, H. Zhu, S. Wang, Y. Wang, D. A. Muller, X. Zhang, *Nat. Nanotechnol.* **2016**.

[41] X. Ling, Y. Lin, Q. Ma, Z. Wang, Y. Song, L. Yu, S. Huang, W. Fang, X. Zhang, A. L. Hsu, Y. Bie, Y.-H. Lee, Y. Zhu, L. Wu, J. Li, P. Jarillo-Herrero, M. Dresselhaus, T. Palacios, J. Kong, *Adv. Mater.* **2016**, *28*, 2322.

[42] M. Mahjouri-Samani, M. Tian, K. Wang, A. Boulesbaa, C. M. Rouleau, A. A. Puretzky, M. A. McGuire, B. R. Srijanto, K. Xiao, G. Eres, G. Duscher, D. B. Geohegan, *ACS Nano* **2014**, *8*, 11567.

[43] K. Wang, B. Huang, M. Tian, F. Ceballos, M.-W. Lin, M. Mahjouri-Samani, A.

Boulesbaa, A. A. Puretzky, C. M. Rouleau, M. Yoon, H. Zhao, K. Xiao, G. Duscher, D. B. Geohegan, *ACS Nano* **2016**, *10*, 6612.

[44] G. Kresse, J. Furthmüller, *Comput. Mater. Sci.* **1996**, *6*, 15.

[45] J. P. Perdew, K. Burke, M. Ernzerhof, *Phys. Rev. Lett.* **1996**, *77*, 3865.

**High conduction hopping behavior induced in transition metal dichalcogenides by percolating defect networks: toward atomically thin circuits**

*Michael G. Stanford[1], Pushpa R. Pudasaini[1], Elisabeth T. Gallmeier[2], Nicholas Cross[1], Liangbo Liang[2], Akinola Oyedele[2], Gerd Duscher[1,3], Masoud Mahjouri-Samani[2], Kai Wang[2], Kai Xiao[2], David B. Geohegan[2], Alex Belianinov[2], Bobby G. Sumpter[2,4], and Philip D. Rack[1,2]\**

1. Department of Materials Science and Engineering, University of Tennessee, Knoxville, Tennessee 37996, United States
2. Center for Nanophase Materials Sciences, Oak Ridge National Laboratory, Oak Ridge, Tennessee 37831, United States
3. Materials Science and Technology Division, Oak Ridge National Laboratory, Oak Ridge, Tennessee 37831, United States
4. Computational Sciences & Engineering Division, Oak Ridge National Laboratory, Oak Ridge, Tennessee 37831, United States

**Dose dependent transfer curves**

**Figure S1a-c** reports the transfer curves for devices before and after they were irradiated with $He^+$. Devices all display p-type characteristics with an on/off of greater than $10^7$.

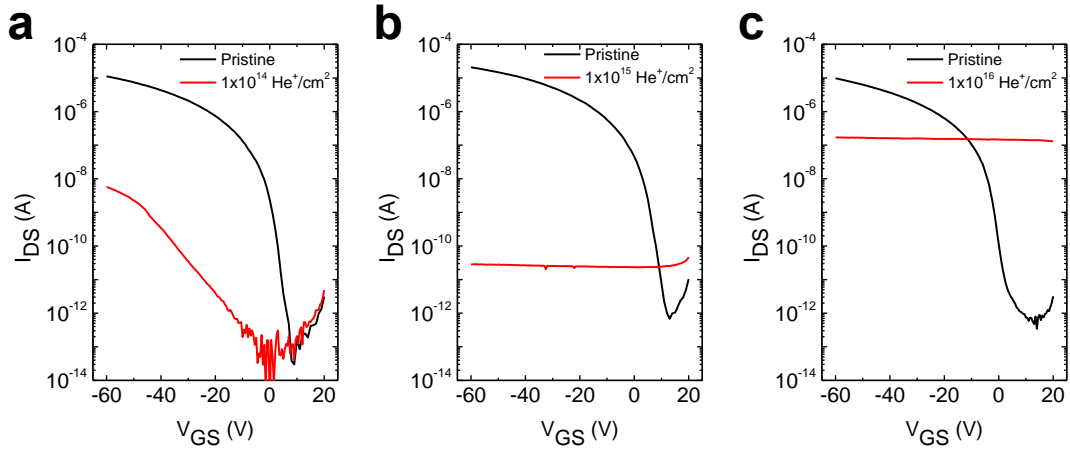

**Figure S2. Transfer curves for device before and after irradiation with the He$^+$ beam. Doses range reported are (a) 1x10$^{14}$ He$^+$/cm$^2$, (b) 1x10$^{15}$ He$^+$/cm$^2$ and (c) 1x10$^{16}$ He$^+$/cm$^2$.**

**Optical properties of WSe$_2$**

**Figure S3a** reports Raman spectra for the WSe$_2$ as a function of He$^+$ irradiation dose. The pristine material exhibits two strong characteristic peaks attributed to the E$^1_{2g}$ and A$_{1g}$ peaks. As the He$^+$ dose increases, the peak intensities decrease, and at a dose of 5×10$^{15}$ He$^+$/cm$^2$ and greater, the E$^1_{2g}$ and A$_{1g}$ peaks are quenched completely, due to defects and lattice distortions generated from He$^+$ irradiation. The photoluminescence (PL) spectra of the irradiated WSe$_2$ are shown in **Figure S3b.** The PL peak at ~1.6 eV is quenched rapidly at doses of 5×10$^{14}$ He$^+$/cm$^2$ and greater, indicating that even relatively low doses introduce a sufficient number defects to quench emission. Similar quenching behavior has been observed for MoS$_2$, where plasma was used to induce defects[1].

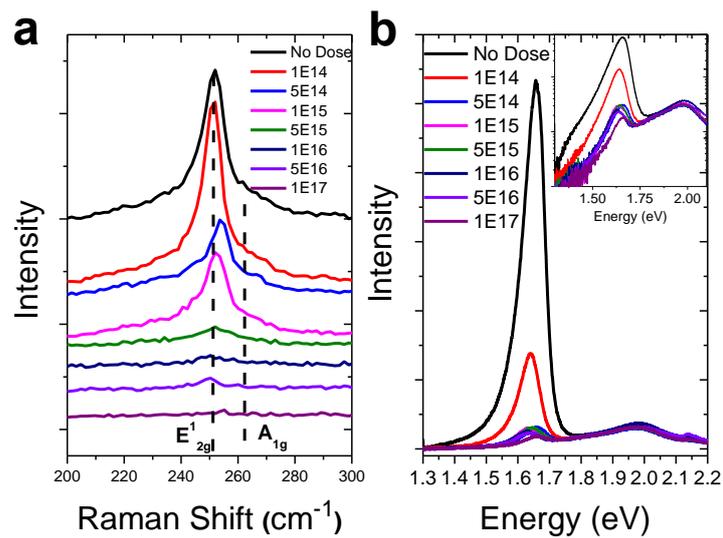

**Figure S3.** (a) Raman spectra for $WSe_2$ as a function of $He^+$ exposure dose. (b) Photoluminescence spectra for $WSe_2$ as a function of $He^+$ exposure dose. Log-scale spectra is inset

**Temperature-dependent current-voltage characteristics**

**Figure S4a** and **Figure S4b** report the temperature dependent source-drain I-V characteristics of a pseudo-metallic device at $V_{GS}$ = 80V and -80V, respectively.

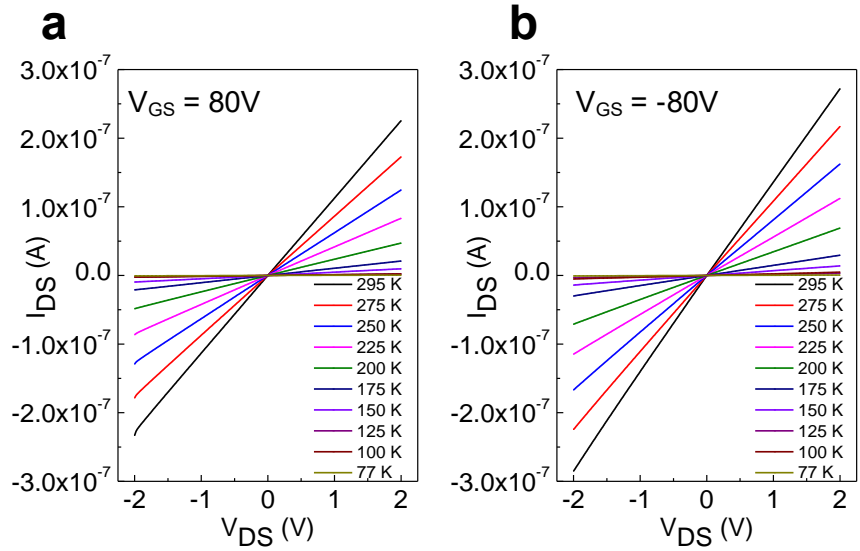

**Figure S4.** Temperature dependent output curves with a $V_{GS}$ of (a) 80 V and (b) -80 V. The device was exposure with a dose of $1.0 \times 10^{16}$ He$^+$/cm$^2$ prior to measurement.

**Electron Energy Loss Spectra (EELS)**

EELS spectra shown in **Figure S5**, contain the selenium $L_{2,3}$ ionization edge at 1436 eV and the W $M_{4,5}$ edge at 1809eV. With increasing dose, the selenium edge decreases while the W edge remains almost constant. Further analysis show that the Se signal is reduced by more than 50% by irradiation with a dose of $1 \times 10^{17}$ He$^+$/cm$^{-1}$. This indicates that the Se concentration is being reduced by interaction with the He$^+$ beam. It is concluded that the He$^+$ preferentially sputters the Se atoms in comparison to W consistent with ion-solid Monte Carlo simulations[2].

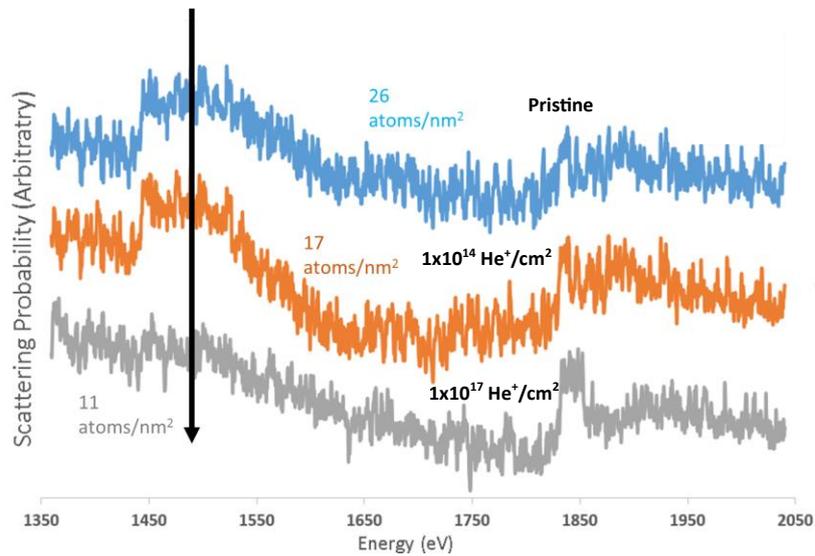

**Figure S5.** EELS spectra showing a decrease in the area under the selenium $L_{2,3}$ edge with increasing $He^+$ dose. This is directly correlated to the areal density of the specific element.

**Inverter device characteristics**

**Figure S6a-c** shows the transfer curves and I-V characteristic for the transistor and resistor portion of the $WS_2$ logic gate. This device was back-gated by $SiO_2$, which has a relatively low dielectric constant. This explains the low voltage gain for the inverter. **Figure S7a,b** show the transfer curves and I-V characteristic for the transistor and resistor portion of the $WSe_2$ logic gate. This device was top-gated with a high dielectric constant ionic liquid.

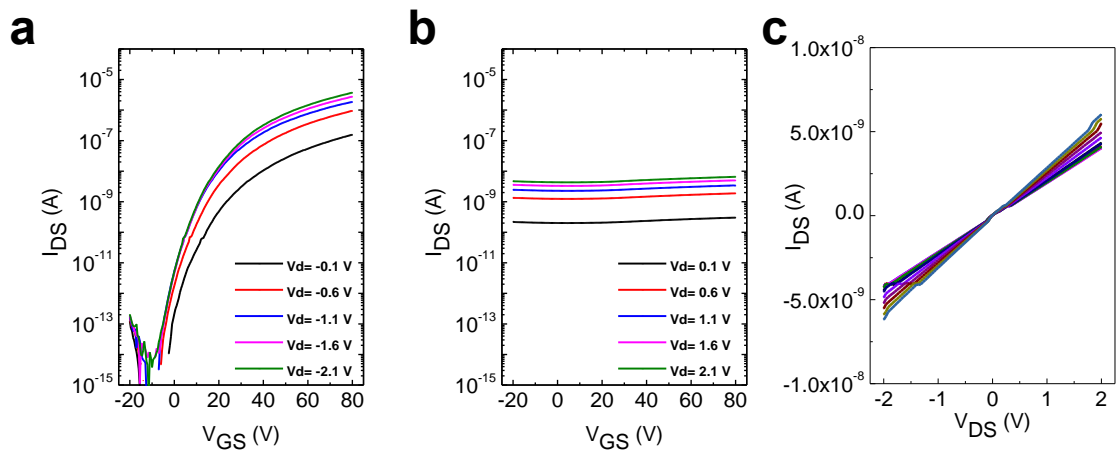

**Figure S6.** (a) Transfer curves for the WS$_2$ transistor which was part of the logic gate. (b) Transfer curves of the metallic-like resistor portion of the inverter which demonstrates no effect of gate modulation. (c) I-V characteristic of the resistor.

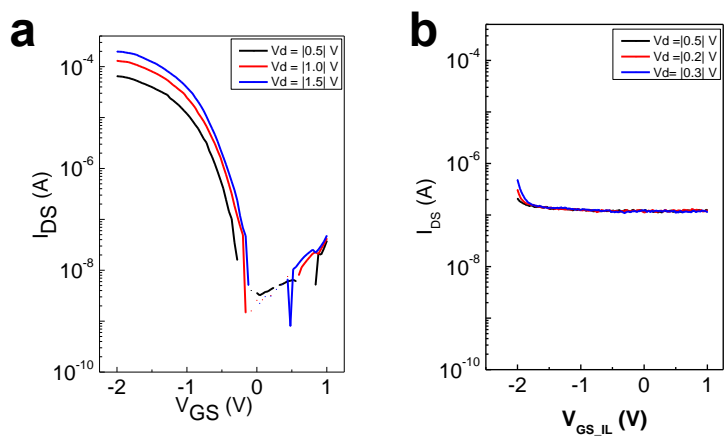

**Figure S7.** (a) Transfer curves for the WSe$_2$ transistor which was part of the logic gate. The device was gated with a high dielectric-constant ionic liquid. (b) Transfer curves of the metallic-like resistor portion of the inverter which demonstrates no effect of gate modulation.

**Details of ionic liquid gating**

1-hexyl-3-methylimidazolium bis(trimetheylsulfonyl)imide ([hmim][Tf2N]) was used as a high dielectric constant ionic liquid (IL) to gate the WSe$_2$ inverter device. A droplet was placed on top of the inverter device in vacuum, and contacted with a prove tip. A schematic of this is shown in **Figure S8.**

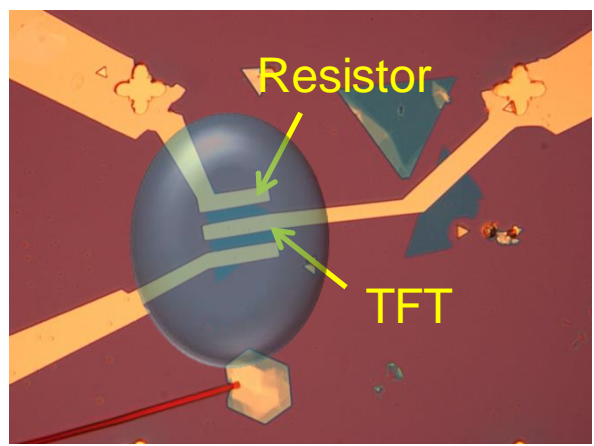

**Figure S8. Schematic of a WSe$_2$ inverter device with IL gating.**

**References**

(1) Kang, N.; Paudel, H. P.; Leuenberger, M. N.; Tetard, L.; Khondaker, S. I. Photoluminescence Quenching in Single-Layer MoS 2 via Oxygen Plasma Treatment. *J. Phys. Chem. C* **2014**, *118*, 21258–21263.

(2) Stanford, M. G.; Pudasaini, P. R.; Belianinov, A.; Cross, N.; Noh, J. H.; Koehler, M. R.; Mandrus, D. G.; Duscher, G.; Rondinone, A. J.; Ivanov, I. N.; *et al.* Focused Helium-Ion Beam Irradiation Effects on Electrical Transport Properties of Few-Layer WSe2: Enabling Nanoscale Direct Write Homo-Junctions. *Sci. Rep.* **2016**, *6*, 27276.